\documentclass[11pt,preprintnumbers,titlepage,nofootinbib]{revtex4-2}
\usepackage[latin9]{inputenc}
\usepackage{amsmath}
\usepackage{amssymb}
\usepackage{graphicx}
\usepackage[unicode=true,
 bookmarks=false,
 breaklinks=false,pdfborder={0 0 1},backref=section,colorlinks=false]
 {hyperref}
\hypersetup{
 colorlinks,linkcolor=blue,citecolor=blue,urlcolor=blue}

\makeatletter

\newcommand{\lyxdot}{.}

\usepackage{wasysym}
\usepackage{array}
\usepackage{multirow}
\usepackage{wasysym}
\usepackage{wasysym}
\usepackage{amsfonts}
\usepackage{subfigure}
\usepackage{cleveref}
\usepackage{listings}
\usepackage{xcolor}
\usepackage{array}
\usepackage{url}
\usepackage{color}
\usepackage{subfigure}
\@ifundefined{textcolor}{}{
\definecolor{BLACK}{gray}{0}
\definecolor{WHITE}{gray}{1}
\definecolor{RED}{rgb}{1,0,0}
\definecolor{GREEN}{rgb}{0,1,0}
\definecolor{BLUE}{rgb}{0,0,1}
\definecolor{CYAN}{cmyk}{1,0,0,0}
\definecolor{MAGENTA}{cmyk}{0,1,0,0}
\definecolor{YELLOW}{cmyk}{0,0,1,0}
\definecolor{ballblue}{rgb}{0.13, 0.67, 0.8}
\definecolor{bleudefrance}{rgb}{0.19, 0.55, 0.91}
\definecolor{blue(ncs)}{rgb}{0.0, 0.53, 0.74}
\definecolor{darkpastelgreen}{rgb}{0.01, 0.75, 0.24}
\definecolor{darkspringgreen}{rgb}{0.09, 0.45, 0.27}
\definecolor{denim}{rgb}{0.08, 0.38, 0.74}
\definecolor{electricviolet}{rgb}{0.56, 0.0, 1.0}
}

\makeatother

\begin{document}
\preprint{CTP-SCU/2023042}
\title{Observations of Orbiting Hot Spots around Scalarized Reissner-Nordström
Black Holes}
\author{Yiqian Chen}
\email{yqchen@stu.scu.edu.cn}

\author{Peng Wang}
\email{pengw@scu.edu.cn}

\author{Haitang Yang}
\email{hyanga@scu.edu.cn}

\affiliation{Center for Theoretical Physics, College of Physics, Sichuan University,
Chengdu, 610064, China}
\begin{abstract}
This paper investigates the observational signatures of hot spots
orbiting scalarized Reissner-Nordström black holes, which have been
reported to possess multiple photon spheres. In contrast to the single-photon
sphere case, hot spots orbiting black holes with two photon spheres
produce additional image tracks in time integrated images capturing
a complete orbit of hot spots. Notably, these newly observed patterns
manifest as a distinct second-highest peak in temporal magnitudes
when observed at low inclination angles. These findings offer promising
observational probes for distinguishing black holes with multiple
photon spheres from their single-photon sphere counterparts.
\end{abstract}
\maketitle
\tableofcontents{}

{}

\section{Introduction}

\label{sec:Introduction}

The Event Horizon Telescope (EHT) collaboration has revolutionized
our understanding of black holes by capturing images of M87{*} and
Sgr A{*}, revealing a striking feature: a bright ring encircling a
dark shadow \cite{Akiyama:2019cqa,Akiyama:2019brx,Akiyama:2019sww,Akiyama:2019bqs,Akiyama:2019fyp,Akiyama:2019eap,Akiyama:2021qum,Akiyama:2021tfw,EventHorizonTelescope:2022xnr,EventHorizonTelescope:2022vjs,EventHorizonTelescope:2022wok,EventHorizonTelescope:2022exc,EventHorizonTelescope:2022urf,EventHorizonTelescope:2022xqj}.
These prominent signatures arise from the strong light deflections
occurring near unstable bound photon orbits, which form photon spheres
in spherically symmetric spacetimes \cite{Synge:1966okc,Bardeen:1972fi,Bardeen:1973tla,Bozza:2009yw}.
These unprecedented observations offer a unique opportunity to directly
test sophisticated theoretical models, such as general relativistic
magnetohydrodynamic simulations, against observational data. This
groundbreaking achievement has ignited a surge of interest in scrutinizing
black hole images illuminated by accreting plasma.

Recent observations and numerical simulations provide compelling evidence
for the formation of hot spots surrounding supermassive black holes.
These highly energized regions are often associated with magnetic
reconnection and flux eruptions within magnetized accretion disks
\cite{Dexter:2020cuv,Scepi:2021xgs,ElMellah:2021tjo}. Notably, recurrent
observations have detected hot spots in close proximity to Sgr A{*}
\cite{Witzel:2020yrp,Michail:2021pgd,GRAVITY:2021hxs}. Additionally,
a notable instance is the detection of an orbiting hot spot within
unresolved light curve data captured at the EHT's observing frequency
\cite{Wielgus:2022heh}. Crucially, the origin of these hot spots
within the compact region near the Innermost Stable Circular Orbit
(ISCO) makes them invaluable for probing black holes in the extreme
gravity regime \cite{abuter2018detection,Wielgus:2022heh}.

Meanwhile, a class of Einstein-Maxwell-scalar (EMS) models has been
introduced to elucidate the formation of hairy black holes \cite{Herdeiro:2018wub}.
These models incorporate non-minimal couplings between the scalar
and Maxwell fields, inducing tachyonic instabilities capable of initiating
spontaneous scalarization. Through fully non-linear numerical simulations,
Herdeiro et al. have demonstrated the transition of Reissner-Nordström
(RN) black holes into scalarized RN black holes \cite{Herdeiro:2018wub}.
This revelation has spurred extensive research within the EMS framework,
exploring various areas such as different non-minimal coupling functions
\cite{Fernandes:2019rez,Fernandes:2019kmh,Blazquez-Salcedo:2020nhs},
massive and self-interacting scalar fields \cite{Zou:2019bpt,Fernandes:2020gay},
horizonless reflecting stars \cite{Peng:2019cmm}, stability analysis
of scalarized black holes \cite{Myung:2018vug,Myung:2019oua,Zou:2020zxq,Myung:2020etf,Mai:2020sac},
higher dimensional scalar-tensor models \cite{Astefanesei:2020qxk},
quasinormal modes of scalarized black holes \cite{Myung:2018jvi,Blazquez-Salcedo:2020jee},
two U(1) fields \cite{Myung:2020dqt}, quasitopological electromagnetism
\cite{Myung:2020ctt}, topology and spacetime structure influences
\cite{Guo:2020zqm}, scalarized black hole solutions in the dS/AdS
spacetime \cite{Brihaye:2019dck,Brihaye:2019gla,Zhang:2021etr,Guo:2021zed,Chen:2023eru},
dynamical scalarization and descalarization \cite{Zhang:2021nnn,Zhang:2022cmu,Jiang:2023yyn}
and rotating scalarized black hole solutions \cite{Guo:2023mda}.

Intriguingly, scalarized RN black holes can harbor multiple photon
spheres outside the event horizon within specific parameter ranges
\cite{Gan:2021pwu}. This unique feature has sparked intensive research
on the optical appearances of various phenomena in their vicinity,
including accretion disks \cite{Gan:2021pwu,Gan:2021xdl,Chen:2023qic},
luminous celestial spheres \cite{Guo:2022muy} and infalling stars
\cite{Chen:2022qrw}. Studies have revealed that an additional photon
sphere significantly amplifies observed accretion disk flux, generates
beat signals in the visibility amplitude, creates triple higher-order
images of a luminous celestial sphere and triggers a cascade of additional
flashes from an infalling star. Furthermore, the presence of multiple
photon spheres raises concerns about spacetime stability due to the
potential for long-lived modes \cite{Cardoso:2014sna,Keir:2014oka,Guo:2021bcw,Guo:2021enm,Guo:2022umh}.
Recent work has shown that these photon spheres can induce superradiance
instabilities for charged scalar perturbations \cite{Guo:2023ivz}.
Moreover, the existence of two photon spheres outside the event horizon
has also been demonstrated in dyonic black holes with a quasi-topological
electromagnetic term \cite{Liu:2019rib,Huang:2021qwe}, black holes
in massive gravity \cite{deRham:2010kj,Dong:2020odp} and wormholes
in the black-bounce spacetime \cite{Tsukamoto:2021caq,Tsukamoto:2021fpp,Tsukamoto:2022vkt}.
For a comprehensive analysis of black holes with multiple photon spheres,
we refer readers to \cite{Guo:2022ghl}.

This paper explores the observational characteristics of hot spots
orbiting scalarized RN black holes, particularly focusing on how the
presence of an additional photon sphere impacts these signatures.
The subsequent sections of this paper are structured as follows: In
Section \ref{sec:SRNBH}, we begin with a concise review of scalarized
RN black hole solutions within the EMS framework, discussing geodesic
motion and gravitational lensing within these spacetimes. Section
\ref{sec:Observation-of-Hot} is devoted to the hot spot model, followed
by an analysis of time integrated images, temporal fluxes and centroids.
Finally, Section \ref{sec:CONCLUSIONS} presents our conclusions.
We adopt the convention $G=c=1$ throughout the paper.

\section{Scalarized RN Black Holes}

\label{sec:SRNBH}

This section first presents a concise review of the scalarized RN
black hole solution within the 4-dimensional EMS model. Following
this, we investigate the properties of photon spheres and ISCOs within
the black hole spacetime.

\subsection{Black Hole Solution}

The EMS model, as outlined in \cite{Herdeiro:2018wub}, combines a
gravity theory with a scalar field $\phi$ and an electromagnetic
field $A_{\mu}$ through the action, 
\begin{equation}
S=\frac{1}{16\pi}\int d^{4}x\sqrt{-g}\left[\mathcal{R}-2\partial_{\mu}\phi\partial^{\mu}\phi-f\left(\phi\right)F_{\mu\nu}F^{\mu\nu}\right],\label{eq:Action}
\end{equation}
where $\mathcal{R}$ is the Ricci scalar, and $F_{\mu\nu}=\partial_{\mu}A_{\nu}-\partial_{\nu}A_{\mu}$
is the electromagnetic field strength tensor. In this EMS model, the
scalar field $\phi$ is non-minimally coupled to the electromagnetic
field $A_{\mu}$ through the coupling function $f\left(\phi\right)$.
For the existence of scalar-free black holes like RN black holes,
the coupling function must satisfy the condition $f^{\prime}\left(0\right)\equiv\left.df\left(\phi\right)/d\phi\right\vert _{\phi=0}=0$
\cite{Herdeiro:2018wub,Fernandes:2019rez}. This study focuses on
the exponential coupling function $f\left(\phi\right)=e^{\alpha\phi^{2}}$
with $\alpha>0$. Within the RN background, the equation of motion
for scalar perturbations $\delta\phi$ is given by 
\begin{equation}
\left(\square-\mu_{\text{eff}}^{2}\right)\delta\phi=0,\label{eq:delta phi}
\end{equation}
where $\mu_{\text{eff}}^{2}=-\alpha Q^{2}/r^{4}$. It is noteworthy
that tachyonic instabilities emerge when the effective mass square
$\mu_{\text{eff}}^{2}$ becomes negative. As demonstrated in \cite{Herdeiro:2018wub,Guo:2021zed},
these instabilities can be sufficiently strong near the event horizon,
inducing the formation of scalarized RN black holes from their scalar-free
counterparts.

To find scalarized RN black hole solutions, we employ the following
ansatz for the metric and electromagnetic field, 
\begin{align}
ds^{2} & =-N\left(r\right)e^{-2\delta\left(r\right)}dt^{2}+\frac{1}{N\left(r\right)}dr^{2}+r^{2}\left(d\theta^{2}+\sin^{2}\theta d\varphi^{2}\right),\nonumber \\
A_{\mu}dx^{\mu} & =\Phi\left(r\right)dt\text{ and}\ \phi=\phi\left(r\right).
\end{align}
In addition, proper boundary conditions are imposed at the event horizon
$r_{h}$ and spatial infinity as 
\begin{align}
N\left(r_{h}\right) & =0\text{, }\delta\left(r_{h}\right)=\delta_{0}\text{, }\phi\left(r_{h}\right)=\phi_{0}\text{, }\Phi\left(r_{h}\right)=0\text{,}\nonumber \\
N\left(\infty\right) & =1\text{, }\delta\left(\infty\right)=0\text{, }\phi\left(\infty\right)=0\text{, }\Phi\left(\infty\right)=\Psi\text{.}\label{eq:infinity condition}
\end{align}
Here, $\delta_{0}$ and $\phi_{0}$ can be used to characterize black
hole solutions, and $\Psi$ is the electrostatic potential. By specifying
$\delta_{0}$ and $\phi_{0}$, we obtain scalarized RN black hole
solutions with a non-trivial scalar field $\phi$ using the shooting
method implemented in $NDSolve$ function of $Wolfram\text{ }\circledR Mathematica$.
Black hole mass $M$ and charge $Q$ are determined from the asymptotic
behavior of the metric functions at infinity 
\begin{align}
N\left(r\right) & =1-\frac{2M}{r}+\cdots,\nonumber \\
\Phi\left(r\right) & =\Psi-\frac{Q}{r}+\cdots.
\end{align}
For simplicity and generality, we express all physical quantities
in units of the black hole mass by setting $M=1$ throughout the paper.

\subsection{Photon Spheres}

The motion of light in scalarized RN black holes is governed by the
geodesic equation, 
\begin{equation}
\frac{d^{2}x^{\mu}}{d\lambda^{2}}+\Gamma_{\rho\sigma}^{\mu}\frac{dx^{\rho}}{d\lambda}\frac{dx^{\sigma}}{d\lambda}=0,\label{eq:geo eq}
\end{equation}
where $\lambda$ is the affine parameter, and $\Gamma_{\rho\sigma}^{\mu}$
represents the Christoffel symbol. For null geodesics characterized
by $ds^{2}=0$, the radial component simplifies to 
\begin{equation}
\frac{e^{-2\delta\left(r\right)}}{L^{2}}\left(\frac{dr}{d\lambda}\right)^{2}+V_{\text{eff}}\left(r\right)=b^{-2},
\end{equation}
where $V_{\text{eff}}\left(r\right)\equiv e^{-2\delta\left(r\right)}N\left(r\right)r^{-2}$
is the effective potential, $b\equiv\left\vert L\right\vert /E$ is
the impact parameter, and $L$ and $E$ are the conserved angular
momentum and energy of photons, respectively. In spherically symmetric
black holes, unstable circular null geodesics constitute a photon
sphere of radius $r_{\text{ph}}$, determined by 
\begin{equation}
V_{\text{eff}}\left(r_{\text{ph}}\right)=b_{\text{ph}}^{-2}\text{, }V^{\prime}\left(r_{\text{ph}}\right)=0\text{, }V^{\prime\prime}\left(r_{\text{ph}}\right)<0,
\end{equation}
where $b_{\text{ph}}$ is the corresponding critical impact parameter.
In other words, a local maximum of the effective potential signifies
the presence of a photon sphere.

\begin{figure}[ptb]
\includegraphics[width=0.45\textwidth]{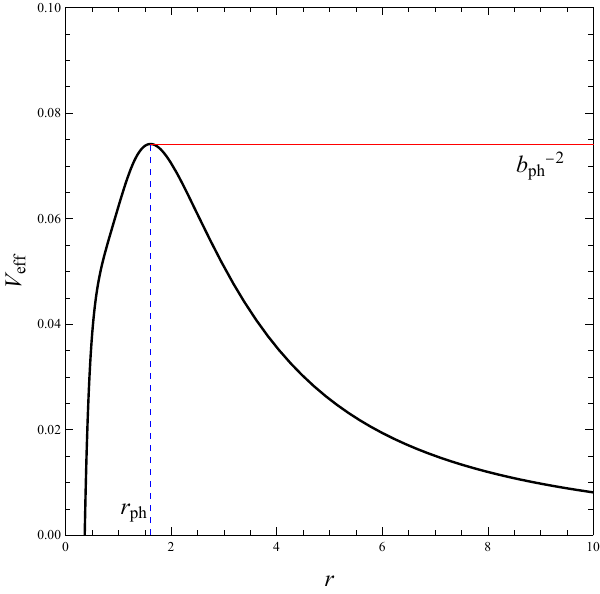} \hspace{20pt}
\includegraphics[width=0.45\textwidth]{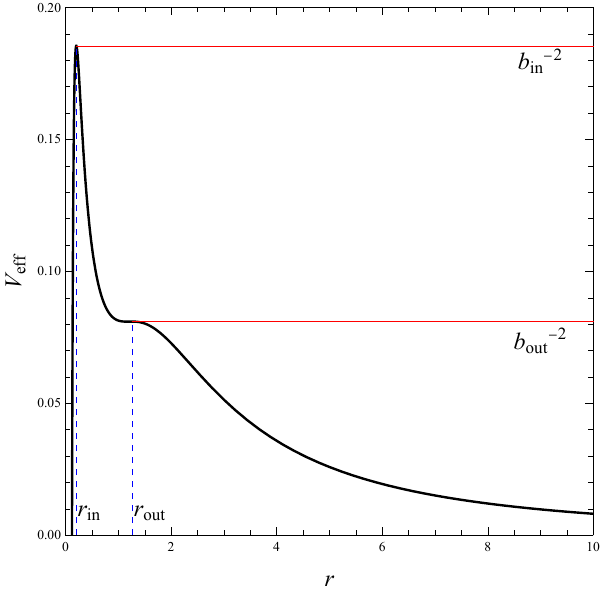} \includegraphics[width=0.43\textwidth]{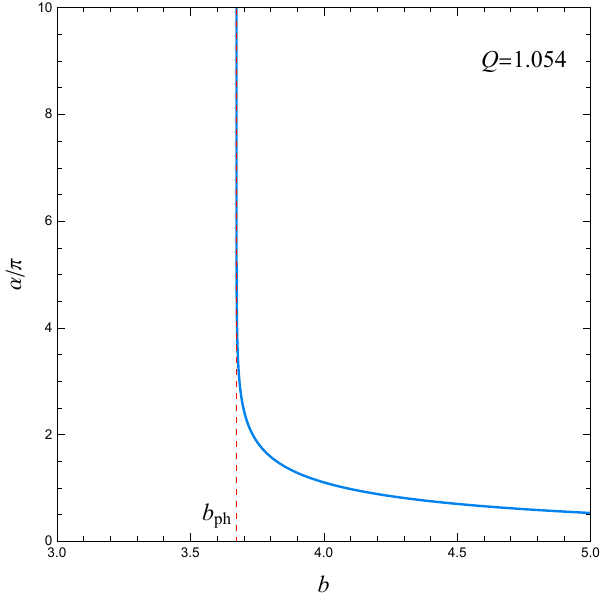}\hspace{35pt}
\includegraphics[width=0.43\textwidth]{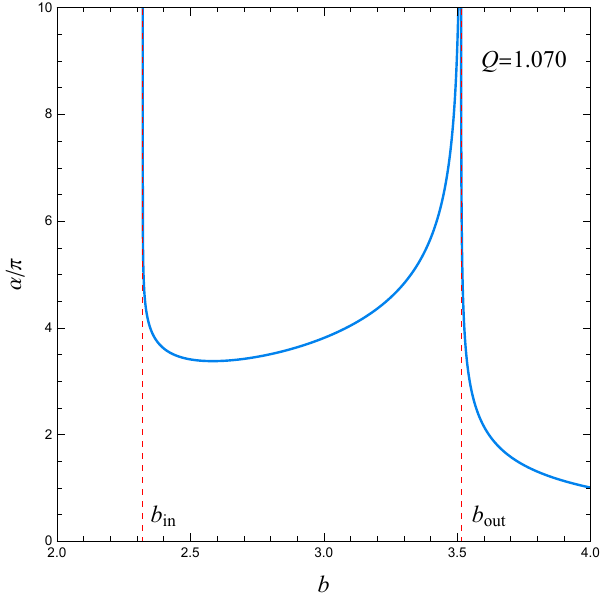}\caption{The effective potential of photons (\textbf{Upper Row}) and the deflection
angle $\alpha$ as a function of the impact parameter $b$ (\textbf{Lower
Row}). \textbf{Left Column: }For $\alpha=0.9$ and $Q=1.054$, the
potential exhibits a single maximum at $r_{\text{ph}}=1.610$, marking
the photon sphere with a critical impact parameter $b_{\text{ph}}=3.673$.
The deflection angle $\alpha$ exhibits a logarithmic divergence at
$b=b_{\text{ph}}$. \textbf{Right Column:} With $\alpha=0.9$ and
$Q=1.070$, the potential reveals two maxima at $r_{\text{in}}=0.204$
and $r_{\text{out}}=1.255$, corresponding to two distinct photon
spheres with critical impact parameters $b_{\text{in}}=2.322$ and
$b_{\text{out}}=3.515$, respectively. Between $b_{\text{in}}$ and
$b_{\text{out}}$, the deflection angle $\alpha$ reaches a minimum
value greater than $3\pi$, indicating that light rays briefly trapped
between the two photon spheres orbit the black hole at least once
before escaping.}
\label{fig:veff}
\end{figure}

Our previous work established that scalarized RN black holes can harbor
either one or two photon spheres outside the event horizon, depending
on the specific values of $\alpha$ and $Q/M$ \cite{Gan:2021pwu}.
FIG. \ref{fig:veff} exemplifies this phenomenon through the effective
potential: the upper-left panel shows a single peak for $\alpha=0.9$
and $Q=1.054$, while the upper-right panel reveals a double-peak
structure for $\alpha=0.9$ and $Q=1.070$. This interesting dependence
led us, in \cite{Guo:2022ghl}, to delineate the region in the $\alpha$-$Q/M$
parameter space where double photon spheres emerge.

To investigate the impact of photon spheres on light deflection, we
focus on the deflection angle $\alpha$ experienced by a light ray
with impact parameter $b$. For rays traveling from and returning
to spatial infinity, the deflection angle is expressed by \cite{Virbhadra:1999nm},
\begin{equation}
\alpha=2\int_{r_{0}}^{\infty}\frac{1}{\sqrt{b^{-2}r^{4}e^{2\delta\left(r\right)}-r^{2}N\left(r\right)}}dr-\pi,
\end{equation}
where $r_{0}$, the turning point, satisfies $V_{\text{eff}}\left(r_{0}\right)=b^{-2}$.
Notably, previous studies have demonstrated that $\alpha$ diverges
logarithmically at critical impact parameters \cite{Virbhadra:1999nm,Bozza:2002zj,Shaikh:2019itn,Chen:2023trn}.
This divergence indicates that light rays undergo multiple circumnavigations
of the black hole in the vicinity of photon spheres, essentially trapped
in their gravitational influence.

The lower-left panel of FIG. \ref{fig:veff} displays the deflection
angle $\alpha$ as a function of $b$ for a scalarized RN black hole
with $\alpha=0.9$ and $Q=1.054$. At large impact parameters, light
rays travel far from the photon sphere, resulting in small deflection.
However, as $b$ approaches the critical value $b_{\text{ph}}$, $\alpha$
diverges logarithmically, reflecting the capture of light rays by
the photon sphere. Note that light rays with $b<b_{\text{ph}}$ plunge
into the event horizon and are excluded from our analysis. The lower-right
panel of FIG. \ref{fig:veff} presents the deflection angle $\alpha$
as a function of $b$ for a scalarized RN black hole with $\alpha=0.9$
and $Q=1.070$. The presence of two photon spheres, marked by dashed
red lines at $b_{\text{in}}$ and $b_{\text{out}}$ (inner and outer,
respectively), distinguishes this case from the single-photon sphere
case in the lower-left panel. For $b>b_{\text{out}}$, $\alpha$ behaves
similarly to the single-photon sphere case, approaching zero with
increasing impact parameter. However, due to the combined gravitational
lensing effects of both photon spheres, light rays with $b_{\text{in}}<b<b_{\text{out}}$
can undergo multiple orbital paths around the black hole, trapped
between the two photon spheres. This orbital confinement manifests
as a minimum in the deflection angle within the range $b_{\text{in}}<b<b_{\text{out}}$.

\subsection{Innermost Stable Circular Orbit}

Finally, we turn our attention to the ISCO of massive particles in
scalarized RN black holes. The ISCO is widely believed to be the origin
of hot spots observed in active galactic nuclei due to synchrotron
radiation emitted by matter orbiting it \cite{Broderick2006,Trippe2007,Hamaus:2008yw}.
Analogously to the null case for photons, the metric condition $ds^{2}=-1/2$
for timelike geodesics leads to the following radial equation, 
\begin{equation}
\frac{e^{-2\delta\left(r\right)}}{L^{2}}\left(\frac{dr}{d\lambda}\right)^{2}+V_{\text{eff}}\left(r\right)=b^{-2},
\end{equation}
where $E$ and $L$ represent the total energy and angular momentum
per unit mass of the orbiting particle, respectively. Here, the effective
potential for massive particles is defined as 
\begin{equation}
V_{\text{eff}}\left(r\right)=e^{-2\delta\left(r\right)}N\left(r\right)\left(\frac{1}{r^{2}}+\frac{1}{L^{2}}\right).
\end{equation}

Consequently, the energy per unit mass $E_{i}$, angular momentum
per unit mass $L_{i}$ and radius $r_{i}$ of the ISCO are identified
by the conditions, 
\begin{equation}
V_{\text{eff}}\left(r_{i}\right)=0\text{, }V_{\text{eff}}^{\prime}\left(r_{i}\right)=0\text{, }V_{\text{eff}}^{\prime\prime}\left(r_{i}\right)=0.
\end{equation}
Once $r_{i}$ is determined, $E_{i}$ and $L_{i}$ can be obtained
from explicit expressions involving the metric functions and their
derivatives at $r_{i}$, 
\begin{align}
E_{i} & =\sqrt{\frac{2N^{2}\left(r_{i}\right)e^{-2\delta\left(r_{i}\right)}}{2N\left(r_{i}\right)-r_{i}N^{\prime}\left(r_{i}\right)+2r_{i}N\left(r_{i}\right)\delta^{\prime}\left(r_{i}\right)}}\text{,}\nonumber \\
\text{ }L_{i} & =\sqrt{\frac{r_{i}^{3}N^{\prime}\left(r_{i}\right)-2r_{i}^{3}\delta^{\prime}\left(r_{i}\right)N\left(r_{i}\right)}{2N\left(r_{i}\right)-r_{i}N^{\prime}\left(r_{i}\right)+2r_{i}N\left(r_{i}\right)\delta^{\prime}\left(r_{i}\right)}}\,\text{.}
\end{align}
For a hot spot orbiting the black hole at the ISCO on the equatorial
plane, its four-velocity takes the form, 
\begin{equation}
v_{e}^{\mu}=\left(\frac{E_{i}}{N\left(r_{i}\right)e^{-2\delta\left(r_{i}\right)}},0,0,\frac{L_{i}}{r_{i}^{2}}\right).
\end{equation}
The corresponding angular velocity and period are $\Omega_{e}=\sqrt{\left[N^{\prime}\left(r_{i}\right)-2\delta^{\prime}\left(r_{i}\right)N\left(r_{i}\right)\right]e^{-2\delta\left(r_{i}\right)}/\left(2r_{i}\right)}$
and $T_{e}=2\pi/\Omega_{e}$, respectively.

\section{Observations of Hot Spot}

\label{sec:Observation-of-Hot}

This section explores observable signatures of hot spots orbiting
scalarized RN black holes. We simplify the analysis by considering
an isotropically emitting hot spot as a sphere. Utilizing the computational
framework described in \cite{Chen:2023uuy,Chen:2023knf}, we place
the observer at $\left(r_{\text{o}},\varphi_{\text{o}}\right)=\left(100,\pi\right)$
with an inclination angle of $\theta_{\text{o}}$, while the hot spot,
with a radius of $0.25$, circles counterclockwise on the ISCO at
$r_{i}$. For optimal precision and efficiency, we employ a $1000\times1000$
pixel grid for each snapshot and generate $200$ snapshots for a full
hot spot orbit. By tracing light rays backward from the observer to
the hot spot, we glean observational information for each image plane
pixel. Specifically, at each time $t_{k}$, each pixel is assigned
an intensity $I_{klm}$, collectively forming lensed images of the
hot spot. Our analysis then focuses on the following image properties,
as outlined in \cite{Hamaus:2008yw,Rosa:2022toh,Rosa:2023qcv},
\begin{itemize}
\item Time integrated image: 
\begin{equation}
\left\langle I\right\rangle _{lm}=\sum\limits _{k}I_{klm},
\end{equation}
summarizing the intensity received at each pixel.
\item Total temporal flux: 
\begin{equation}
F_{k}=\sum\limits _{l}\sum\limits _{m}\Delta\Omega I_{klm},
\end{equation}
representing the total intensity received at time $t_{k}$. Here,
$\Delta\Omega$ denotes the solid angle per pixel.
\item Temporal magnitude: 
\begin{equation}
m_{k}=-2.5\lg\left(\frac{F_{k}}{\min\left(F_{k}\right)}\right),
\end{equation}
quantifying the relative brightness of each snapshot.
\item Temporal centroid: 
\begin{equation}
\overrightarrow{c_{k}}=F_{k}^{-1}\sum\limits _{l}\sum\limits _{m}\Delta\Omega I_{klm}\overrightarrow{r_{lm}},
\end{equation}
indicating the center of intensity distribution in each snapshot.
Here, $\overrightarrow{r_{lm}}$ represents the position relative
to the image center. 
\end{itemize}

\subsection{Integrated Images}

\begin{figure}[ptb]
\includegraphics[width=0.45\textwidth]{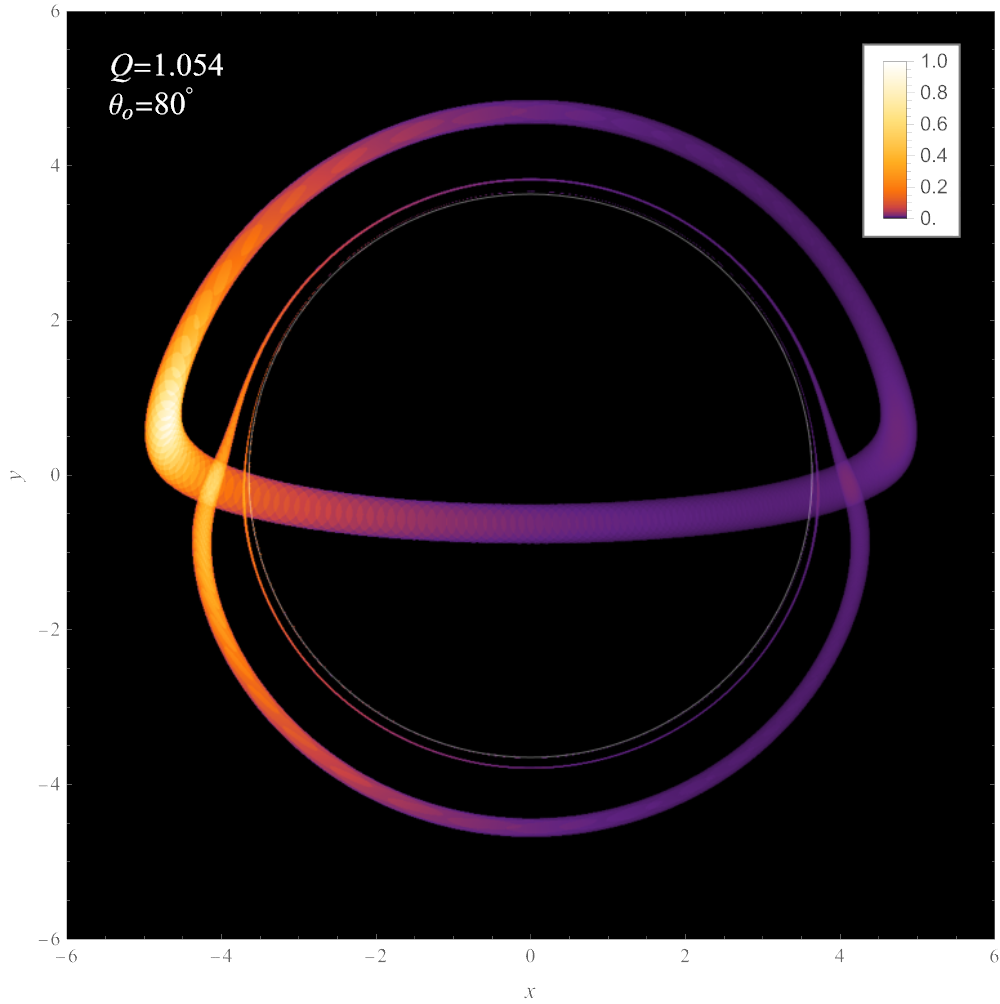}
\hspace{10pt} \includegraphics[width=0.45\textwidth]{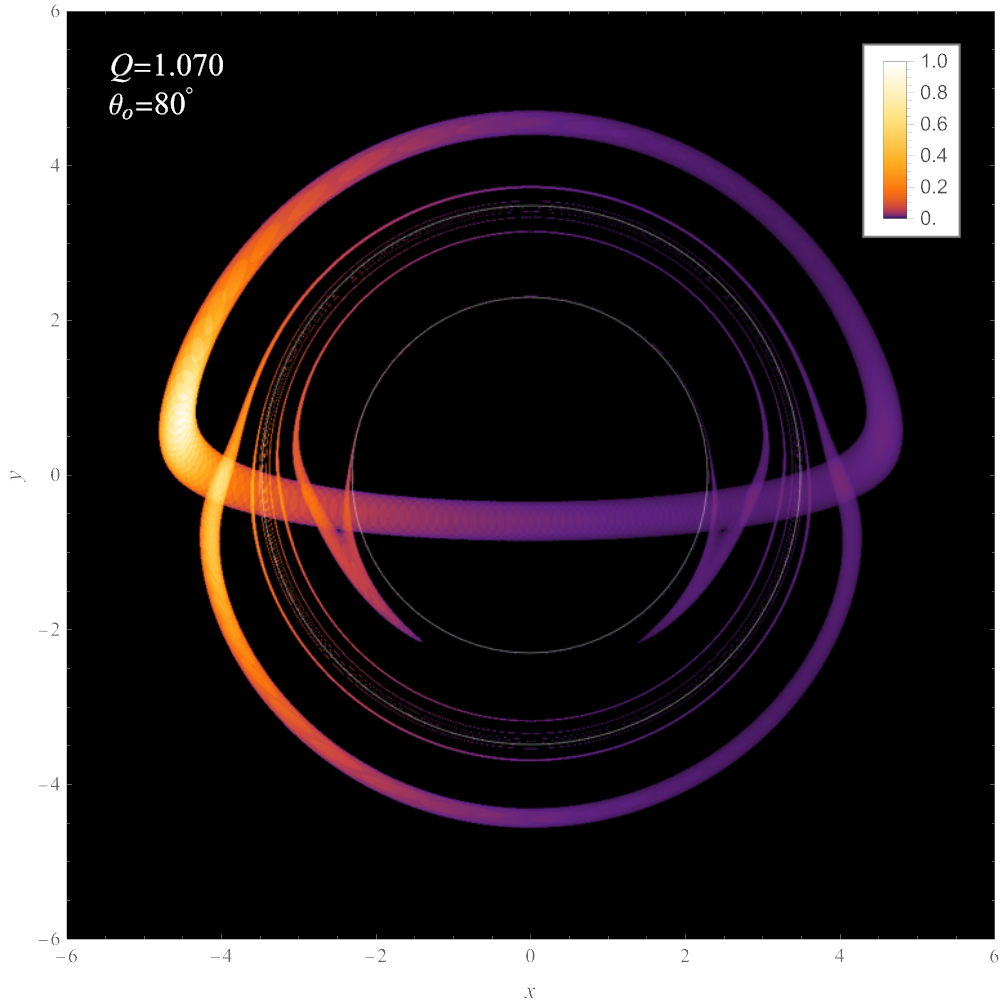}
\vspace{0pt}
 \includegraphics[width=0.45\textwidth]{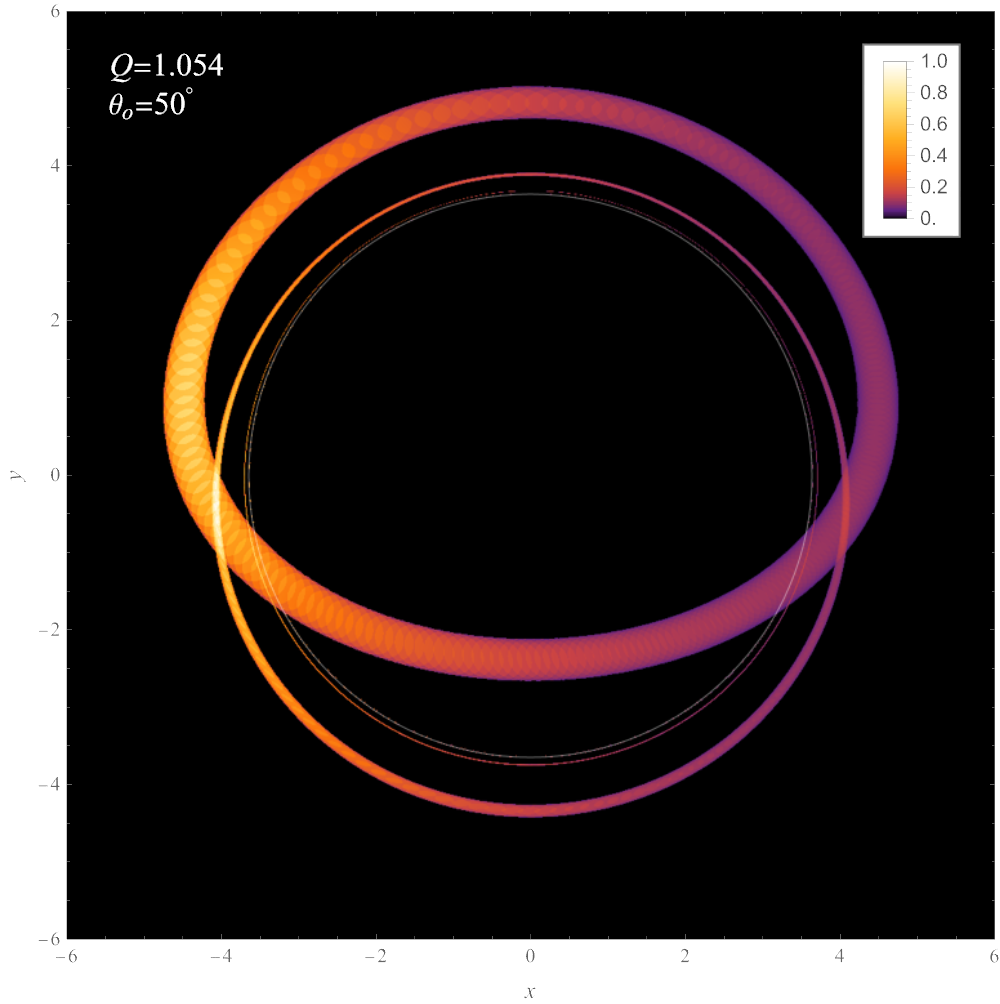}
\hspace{10pt} \includegraphics[width=0.45\textwidth]{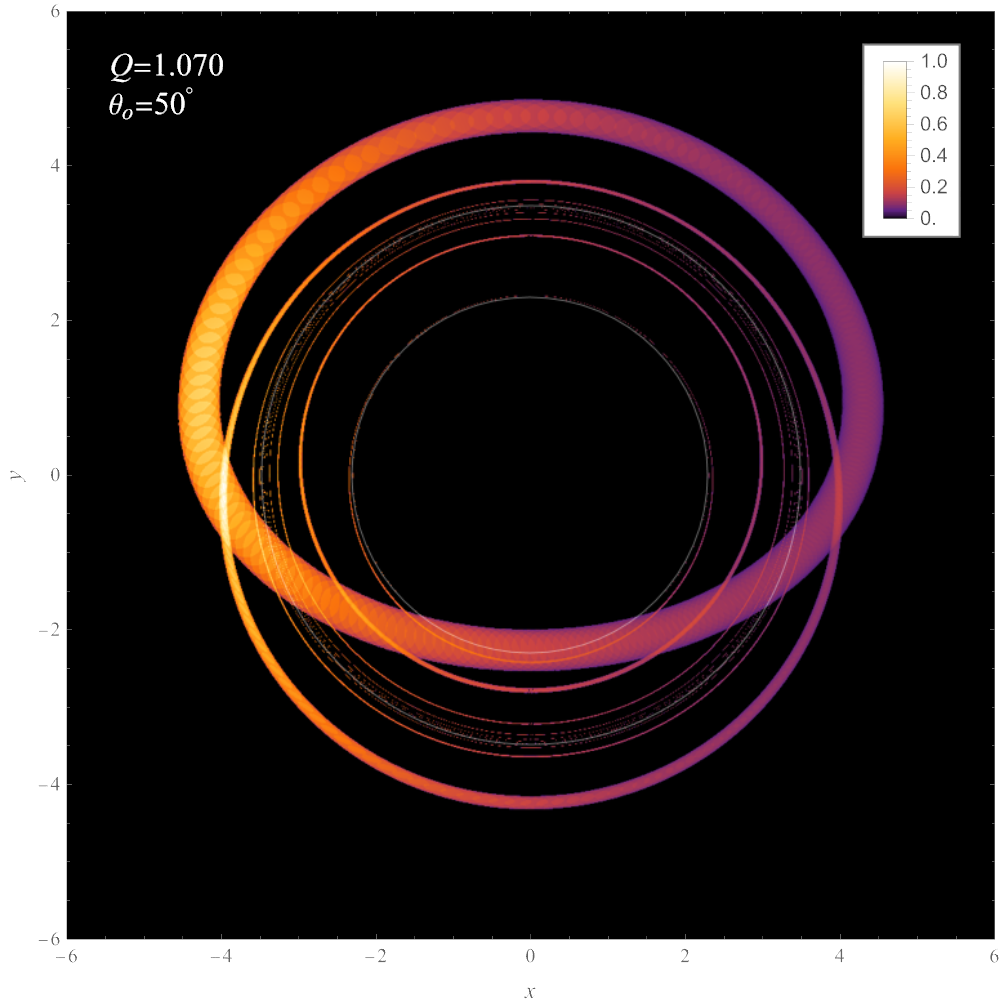}\caption{Time integrated images for a complete orbit of the hot spot, captured
from an observational inclination angle of $\theta_{\text{o}}=80^{\circ}$
(\textbf{Upper Row}) and $50^{\circ}$ (\textbf{Lower Row}). The critical
curves, outlined in white, are traced by light rays escaping the photon
spheres. All intensity values are normalized to their maximum value.
\textbf{Left Column:} Single-photon sphere case with $\alpha=0.9$
and $Q=1.054$. The images highlight the primary, secondary and tertiary
lensed image tracks positioned beyond the critical curve, stemming
from the $n=0^{>}$, $1^{>}$ and $2^{>}$ emissions of the hot spot,
respectively. \textbf{Right Column:} Double-photon sphere case with
$\alpha=0.9$ and $Q=1.070$. The images unveil three image tracks
outside the outer critical curve, alongside three additional tracks
between the two critical curves. The two innermost tracks between
the critical curves originate from $n=4^{<>}$ light rays. For $\theta_{\text{o}}=80^{\circ}$,
they become incomplete and merge into Y-shaped patterns. The remaining
track arises from $n=5^{<>}$ light rays. }
\label{TI-images}
\end{figure}

FIG. \ref{TI-images} displays the time integrated images of scalarized
RN black holes with one or two photon spheres, viewed at inclination
angles of $80^{\circ}$ and $50^{\circ}$. White lines mark critical
curves traced by light rays escaping the photon spheres. While the
single-photon sphere case shows three prominent image tracks outside
the critical curve, the double-photon sphere case exhibits additional
tracks between the two critical curves. To decipher the origin of
these tracks, we use a numerical count $n$ signifying the number
of equatorial plane crossings during a light ray's journey, characterizing
its path and resulting image track. Note that, in double-photon sphere
cases, photons can orbit between the inner and outer spheres. Therefore,
superscripts $>$ and $<>$ denote light rays traveling outside the
(outer) photon sphere and orbiting between the inner and outer photon
spheres, respectively.

Focusing on an observer at $\theta_{\text{o}}=80^{\circ}$, the upper-left
panel of FIG. \ref{TI-images} presents the time integrated image
of the hot spot for a single-photon sphere scalarized RN black hole
with $\alpha=0.9$ and $Q=1.054$. A striking brightness asymmetry,
arising from the Doppler effect, is immediately apparent. Similar
to the Schwarzschild black hole case (FIG. 2 in \cite{Chen:2023knf}),
the primary image ($n=0^{>}$) forms a closed semicircular track,
with upper and lower segments representing the hot spot appearing
behind and in front of the black hole. The smaller and dimmer track
represents the secondary image ($n=1^{>}$) with its scarcely visible
upper segment corresponding to the front-positioned hot spot and the
lower segment to the back-positioned one. Unlike the Schwarzschild
case, a faint and more circular tertiary track ($n=2^{>}$) appears
near the critical curve due to the flatter potential peak of this
specific configuration.

The upper-right panel of FIG. \ref{TI-images} depicts the time integrated
image for the double-photon sphere case with $\alpha=0.9$ and $Q=1.070$.
Familiar tracks for $n=0^{>}$, $1^{>}$ and $2^{>}$ reside outside
the outermost critical curve, similar to the single-photon sphere
case. However, fascinatingly, additional tracks emerge between the
two critical curves, originating from light rays orbiting between
the inner and outer photon spheres. As revealed in FIG. \ref{fig:veff},
these orbits have a minimum value of the deflection angle $\alpha$,
setting a lower bound on $n$ of $n\geq4$. Importantly, two distinct
impact parameters can lead to the same $\alpha$ above the minimum,
resulting in two separate $n=4^{<>}$ tracks. Due to this minimum
$\alpha$ value, the two $n=4^{<>}$ tracks are incomplete and merge
into Y-shaped patterns between the critical curves. A fainter $n=5^{<>}$
track appears outside the $n=4^{<>}$ tracks, and higher-order images
become increasingly less luminous and hug the critical curves.

The lower panels of FIG. \ref{TI-images} present time integrated
images obtained at an observation inclination of $\theta_{\text{o}}=50^{\circ}$,
revealing similarities to the $\theta_{\text{o}}=80^{\circ}$ cases.
Despite these similarities, an observer positioned at a lower inclination
angle perceives a reduced degree of brightness asymmetry, accompanied
by a more circular appearance of the image tracks. Moreover, for the
double-photon sphere case, $n=4^{<>}$ light rays now form two complete
tracks, as the deflection angle is unable to reach its minimum value
at this observation inclination.

\subsection{Temporal Fluxes and Centorids}

\begin{figure}[ptb]
\includegraphics[width=0.45\textwidth]{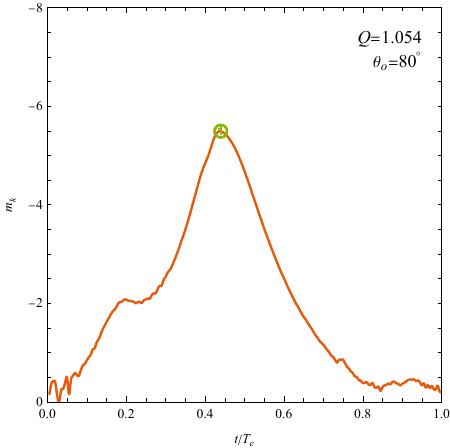} \hspace{10pt}
\includegraphics[width=0.45\textwidth]{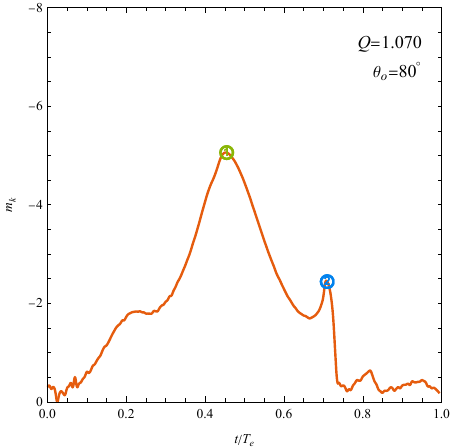} 

\includegraphics[width=0.46\textwidth]{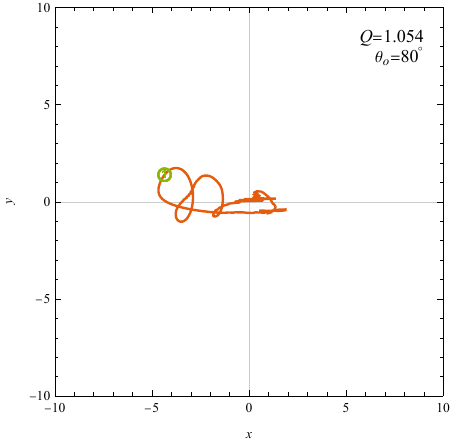} \hspace{8pt}
\includegraphics[width=0.46\textwidth]{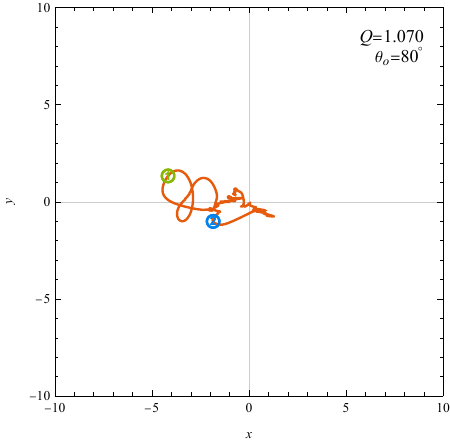}\caption{Temporal magnitudes $m_{k}$ (\textbf{Upper Row}) and centroids $c_{k}$
(\textbf{Lower Row}) as a function of $t/T_{e}$. The inclination
is $\theta_{o}=80^{\circ}$. \textbf{Left Column:} Single-photon sphere
case with $\alpha=0.9$ and $Q=1.054$. The green circle \textcolor{darkpastelgreen}{\textcircled{1}}
marks the magnitude peak dominated by the primary image. \textbf{Right
Column:} Double-photon sphere case with $\alpha=0.9$ and $Q=1.070$.
The green circle \textcolor{darkpastelgreen}{\textcircled{1}}
and blue circle \textcolor{denim}{\textcircled{2}} mark the highest
and second-highest peaks, respectively. The additional peak compared
to the single-photon sphere case arises from light rays orbiting between
the two photon spheres.}
\label{Fig: mkck-theta80}
\end{figure}

\begin{figure}[ptb]
\includegraphics[width=0.45\textwidth]{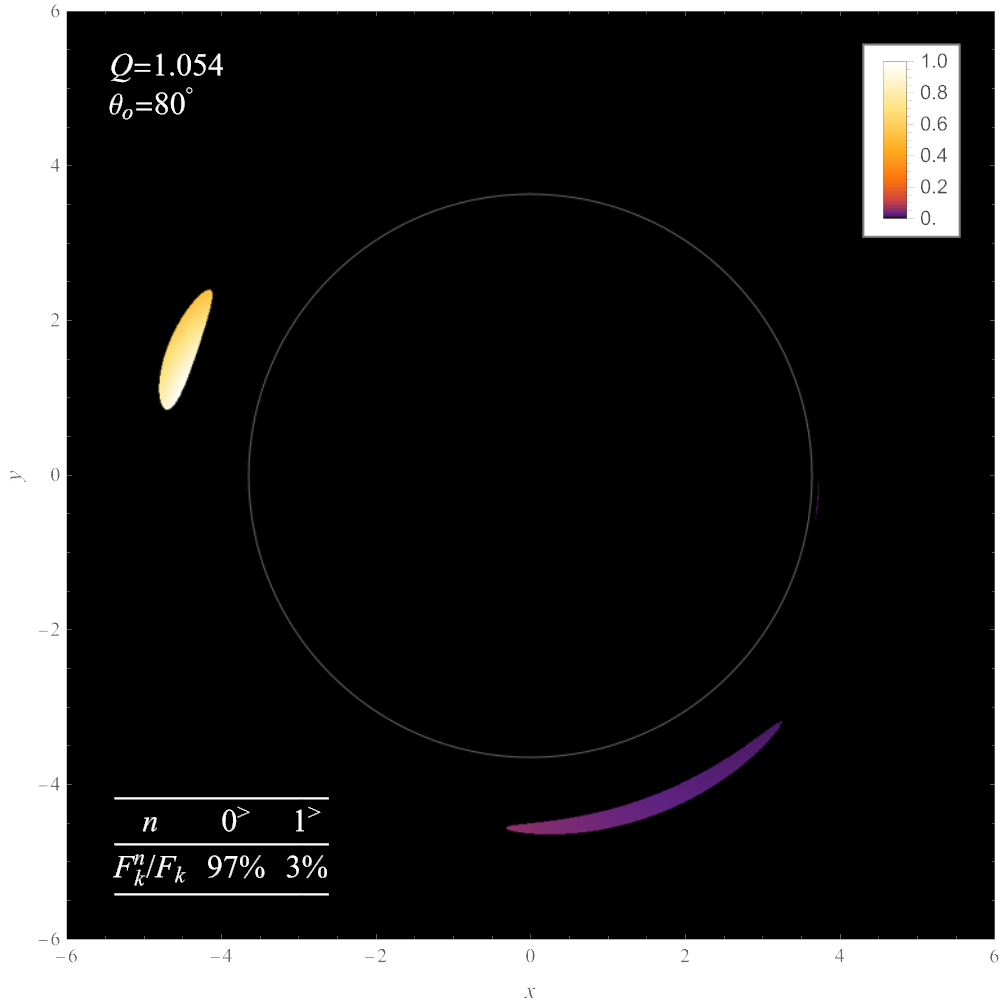}
\hspace{10pt} \includegraphics[width=0.45\textwidth]{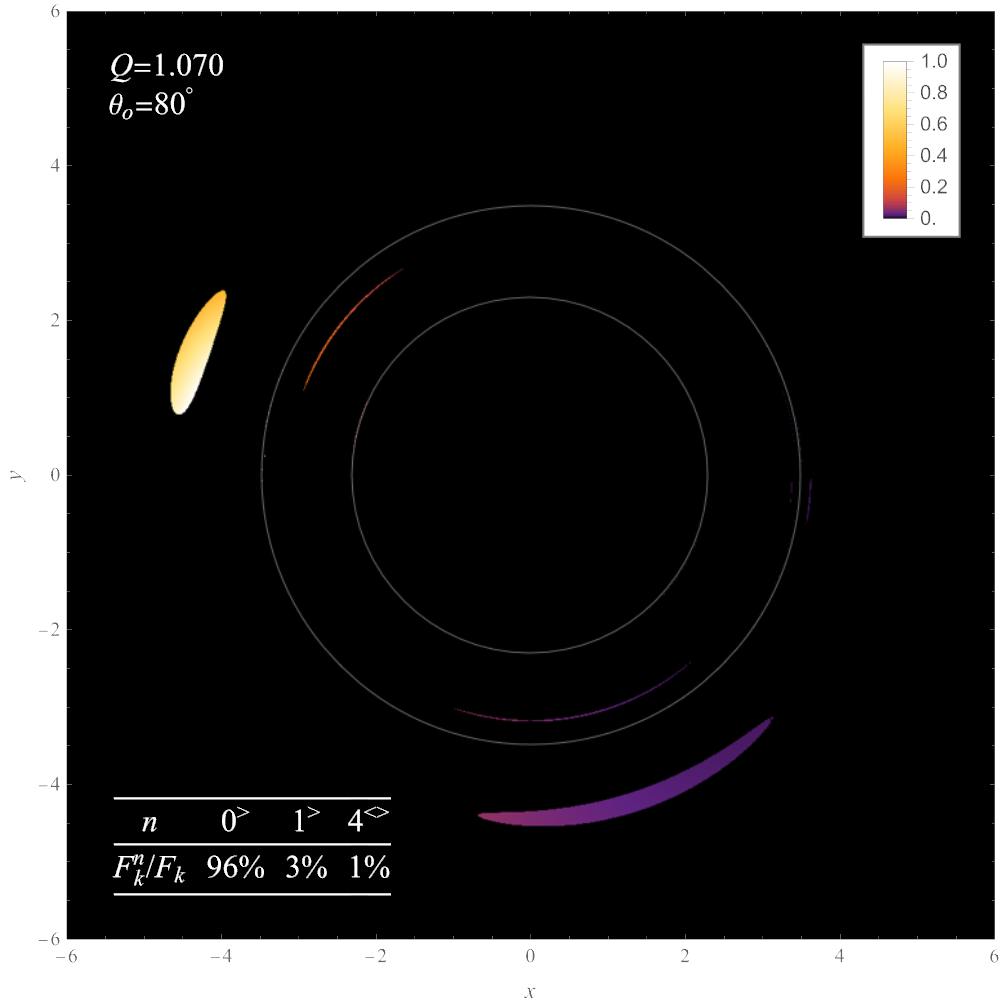} 

\hspace{0.49\textwidth}\includegraphics[width=0.45\textwidth]{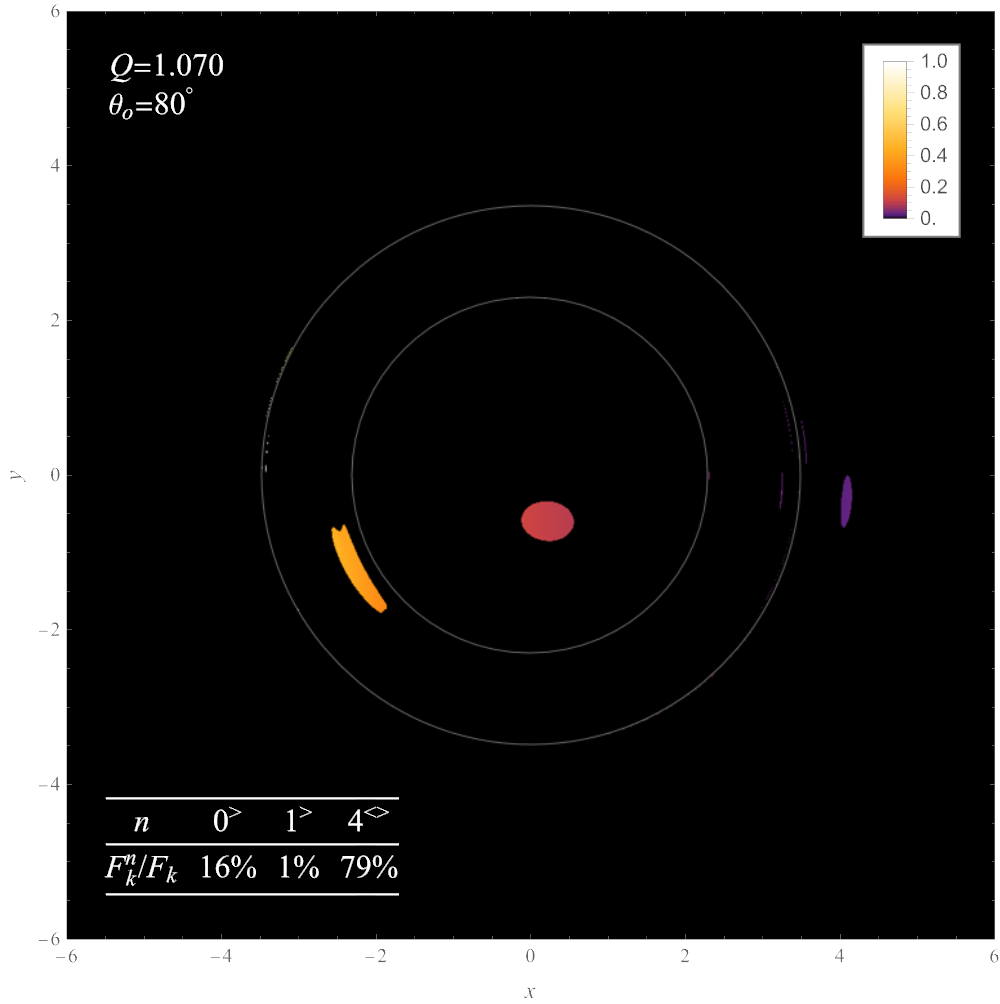}\caption{Snapshots for the single-photon sphere (\textbf{Left Column}) and
double-photon sphere (\textbf{Right Column}) cases, captured at the
maximum temporal magnitude. The upper and lower rows depict the snapshots
corresponding to the highest and second-highest peaks, respectively.
The relative contribution of the $n^{\text{th}}$-order image to the
total flux $F_{k}$ is given by $F_{k}^{n}/F_{k}$, where $F_{k}^{n}$
denotes the temporal flux of the $n^{\text{th}}$-order image at $t=t_{k}$.
The second-highest peak in the double-photon sphere case arises primarily
from the $n=4^{<>}$ image.}
\label{Fig: snapshots-theta80}
\end{figure}

FIG. \ref{Fig: mkck-theta80} illustrates the temporal magnitudes
$m_{k}$ and centroids $c_{k}$ as a function of time observed at
$\theta_{\text{o}}=80^{\circ}$ for scalarized RN black holes with
a single photon sphere or double photon spheres. In the case of a
single photon sphere, a prominent peak, labeled \textcolor{darkpastelgreen}{\textcircled{1}},
is evident in the temporal magnitude plot shown in the upper-left
panel of FIG. \ref{Fig: mkck-theta80}. The corresponding snapshot,
highlighting the contributions from various images with differing
values of $n$ as tabulated, is displayed in the upper-left panel
of FIG. \ref{Fig: snapshots-theta80}. Notably, the flux at this peak
is predominantly contributed by the primary images with $n=0^{>}$,
originating from the hot spot located near the leftmost portion of
its orbit. This observation aligns with expectations, as the hot spot
moves closer to the observer on the left side of the field of view,
leading to a pronounced increase in the observed light frequency due
to the Doppler effect. The lower-left panel of FIG. \ref{Fig: mkck-theta80}
illustrates the centroid, which mainly resides in the left half of
the image plane. When the primary image experiences a Doppler-induced
flux reduction in the right half of the plane, the secondary image
can significantly shift the centroid towards the left. Additionally,
higher-order images introduce numerical noise in regions of low flux,
impacting both magnitudes and centroids.

In the double-photon sphere case, two distinct peaks are visible in
the temporal magnitude plot, labeled \textcolor{darkpastelgreen}{\textcircled{1}}
and \textcolor{denim}{\textcircled{2}} in the upper-right panel
of FIG. \ref{Fig: mkck-theta80}, respectively. The snapshot corresponding
to the highest peak, displayed in the upper-right panel of FIG. \ref{Fig: snapshots-theta80},
bears similarities to the single-photon sphere case. Notably, the
snapshot at the second-highest peak, presented in the lower-right
panel of FIG. \ref{Fig: snapshots-theta80}, unveils a departure from
the exclusive dominance of primary images. This shift occurs as the
primary image undergoes a phase of diminished flux due to the hot
spot's movement away from the observer. Consequently, the $n=4^{<>}$
image emerges as a significant contributor to the overall flux, resulting
in a pronounced local magnitude peak. Furthermore, as demonstrated
in the lower-right panel of FIG. \ref{Fig: mkck-theta80}, the influence
of higher-order images between the two critical curves tends to displace
the centroid further towards the left compared to the single-photon
sphere case.

\begin{figure}[ptb]
\includegraphics[width=0.45\textwidth]{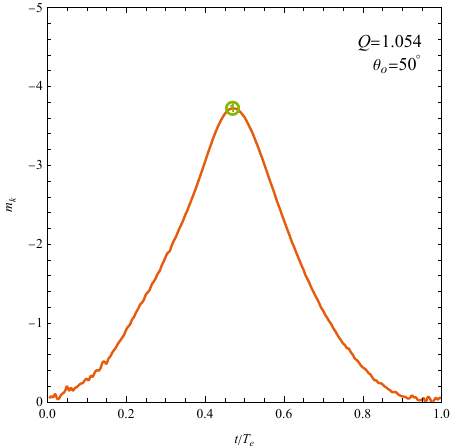} \hspace{10pt}
\includegraphics[width=0.45\textwidth]{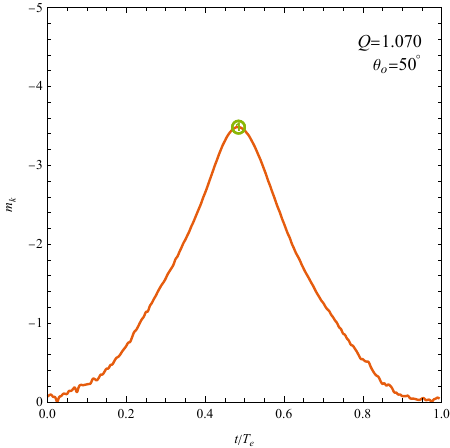} 

\includegraphics[width=0.46\textwidth]{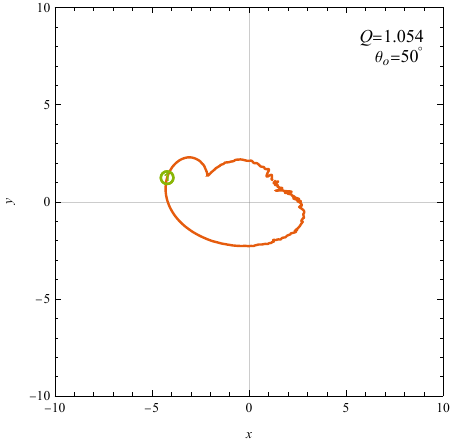} \hspace{8pt}
\includegraphics[width=0.46\textwidth]{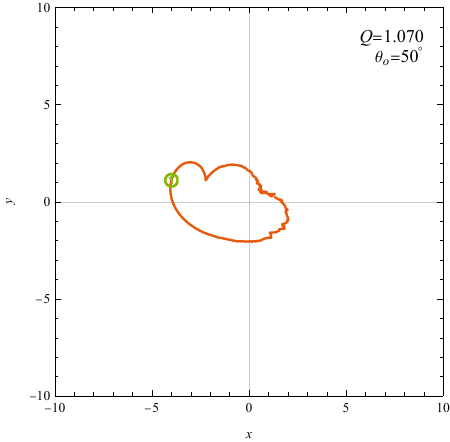}\caption{Temporal magnitudes $m_{k}$ (\textbf{Upper Row}) and centroids $c_{k}$
(\textbf{Lower Row}) plotted against $t/T_{e}$ for the single-photon
sphere (\textbf{Left Column}) and double-photon sphere (\textbf{Right
Column}) cases at an inclination of $\theta_{o}=50^{\circ}$. Due
to the reduced influence of the Doppler effect at lower inclinations,
both cases display a single peak in the temporal magnitudes, and the
centroids are shifted towards the center.}
\label{Fig: mkck-thet50}
\end{figure}

Figure \ref{Fig: mkck-thet50} displays the temporal magnitudes and
centroids for an inclination angle of $\theta_{o}=50^{\circ}$. In
contrast to the $\theta_{o}=80^{\circ}$ case, only a single peak
is visible in the temporal magnitudes for $\theta_{o}=50^{\circ}$.
This difference arises due to the decreased impact of the Doppler
effect at the lower inclination angle. Consequently, the flux becomes
less frequency-dependent, enabling the primary image to dominate the
total flux contribution for most of the time. Moreover, the influence
of higher-order images on the centroids diminishes, causing it to
shift toward the center of the primary image's orbit.

\section{Conclusions}

\label{sec:CONCLUSIONS}

In this paper, we have examined the observable behavior of hot spots
orbiting scalarized RN black holes along the ISCOs. Intriguingly,
depending on the parameters of the black hole, a scalarized RN black
hole may possess either one or two photon spheres \cite{Gan:2021xdl}.
Given the substantial influence of photon spheres on black hole imaging,
our observations have revealed distinctive characteristics between
the single-photon sphere and double-photon sphere cases.

In the single-photon sphere case, the primary observational features
align closely with those observed in the Schwarzschild black hole
case \cite{Chen:2023knf}. Specifically, the primary image with $n=0^{>}$
traces a closed semicircular track, culminating in a pronounced peak
within the temporal flux. Conversely, in the case of double-photon
spheres, photons have the capacity to orbit between the two photon
spheres multiple times, resulting in the generation of additional
hot spot images situated between the two critical curves. Consequently,
when viewed at lower inclinations, these additional images contribute
to a distinct secondary peak in the temporal flux.

Through the analysis of these image characteristics, we are able to
gain deeper insights into the optical manifestations of hot spots
in proximity to black holes. This understanding paves the way for
the discrimination between black holes harboring a single photon sphere
and those possessing multiple photon spheres. The advent of next-generation
Very Long Baseline Interferometry holds promising potential for leveraging
our findings as a means of exploring black holes with multiple photon
spheres.
\begin{acknowledgments}
We are grateful to Guangzhou Guo and Tianshu Wu for useful discussions
and valuable comments. This work is supported in part by NSFC (Grant
No. 12275183, 12275184 and 11875196). 
\end{acknowledgments}

 \bibliographystyle{unsrturl}
\bibliography{ref}

\end{document}